\documentclass{iau}
\usepackage{graphicx}

\title[Observations of G2] %% give here short title %%
{Observations of the gas cloud G2 in the Galactic Center}

\author[Gillessen et al.]   %% give here short author list %%
{
Stefan Gillessen$^1$,
Reinhard Genzel$^{1,2}$,
Tobias K. Fritz$^1$,\\
Frank Eisenhauer$^1$,
Oliver Pfuhl$^1$,
Thomas Ott$^1$, 
Andreas Burkert$^{1,3}$, 
Marc Schartmann$^{1,3}$, 
Alessandro Ballone$^{1,3}$
}

\affiliation{
$^1$Max-Planck-Institute for Extraterrestrial Physics,\\ Giessenbachstrasse, 85748 Garching, Germany \\ 
$^2$ Departments of Physics and Astronomy, Le Conte Hall,\\ University of California, 94720 Berkeley, USA\\ 
$^3$ Universit\"atssternwarte der Ludwig-Maximilians-Universit\"at, \\ Scheinerstr. 1, D-81679 M\"unchen, Germany
}

\pubyear{2013}
\volume{303}  %% insert here IAU Symposium No.
\pagerange{119--126}
\jname{The Galactic Center: feeding \& feedback in a normal gal. nucleus}
\editors{A.C. Editor, B.D. Editor \& C.E. Editor, eds.}
\begin{document}

\maketitle

\begin{abstract}
In 2011, we discovered a compact gas cloud (''G2'') with roughly three Earth
masses that is falling on a near-radial orbit toward the massive
black hole in the Galactic Center. The orbit is well constrained and
pericenter passage is predicted for early 2014. Our data beautifully
show that G2 gets tidally sheared apart due to the massive black hole's force. During the next months, we expect that in addition to the tidal effects, hydrodynamics get important, when G2 collides with the hot ambient gas around Sgr A*. Simulations show that ultimately, the cloud's material might fall into the massive black hole. Predictions for the accretion rate and luminosity evolution, however, are very difficult due to the many unknowns. Nevertheless, this might be
a unique opportunity in the next years to observe how gas feeds a massive black hole in a galactic nucleus.
\end{abstract}

\firstsection % if your document starts with a section,
              % remove some space above using this command.
\section{Introduction}

In 2011, we made a surprising discovery: Our long-term Very Large Telescope-based, near-infrared observations of the Galactic Center (GC) showed a small gas cloud (G2) falling on a nearly radial orbit onto Sgr A* (\cite[Gillessen et al. 2012]{gil12}). In particular, we detected a temporally increasing velocity shear of G2's line emission in deep integral field spectroscopy data. This is the unambiguous sign of the massive black hole's (MBH) tidal field. We have followed up the evolution with similar observations in 2012 (\cite[Gillessen et al. 2013a]{gil13a}) and 2013 (\cite[Gillessen et al. 2013b]{gil13b}), spectacularly showing the onset of the disruption of G2. The case caught the immediate attention of a broad audience, since this might constitute the unique opportunity to watch in real-time, how a MBH is getting fed. 

Here, we summarize our observations, constraining its orbit and properties. We present a sequence of position-velocity diagrams, showing the tidal disruption, which can be well-described by a simple test particle model. Adopting a gas cloud model, hydrodynamic simulations predict, that this description remains a good approximation until pericenter. Afterwards the further evolution is dominated by hydrodynamics (\cite[Schartmann et al. 2012]{sch12}, \cite[Anninos et al. 2012]{ann12}). The motion remains of course Keplerian throughout for models with a stellar source inside. We try to summarize what is known about the nature of G2, i.e. its origin, and finally present ideas, what future observations across all wavebands might be able to detect.

\section{The orbit of G2}
G2 caught our attention as a fast moving L-band ($\approx 4\,\mu$m) source that apparently was on a curved trajectory towards Sgr~A*. It did not show a K-band counterpart, but we were able to see strong Brackett-$\gamma$ emission spatially coincident in our SINFONI (\cite[Eisenhauer et al. 2003]{eis03}, \cite[Bonnet et al. 2004]{bon04}) data. The line position changed consistently over the years. In total, we obtained eight dynamical quantities: the position on sky (2), the proper motion (2), the acceleration (1), the radial velocity $v_\mathrm{LSR}$ (1), its change with time (1), and even a second derivative of $v_\mathrm{LSR}$ (1). An orbit has six free parameters, so it was non-trivial that we were able to actually find an orbit describing the data. In turn, it means that it is very probable that astrometry and $v_\mathrm{LSR}$ data belong to the same object.
The initial orbit estimate had an estimated time of pericenter passage of around mid 2013 (\cite[Gillessen et al. 2012]{gil12}). 

\cite{phi13} showed convincingly that there is a systematic offset in the L-band positions compared to the positions as derived from Brackett-$\gamma$. This is probably due to the underlying, faint and spatially variable dust emission. The main difference is that the pericenter date of their orbit is shifted by a few months to early 2014. These authors used the Keck-based OSIRIS instrument (\cite[Larkin et al. 2006]{lar06}). It uses a lenslet array to achieve the integral field spectroscopy, as compared to SINFONI that uses an image slicer. The former intrinsically has a cleaner astrometric performance. Nevertheless, we were able to reproduce the same orbit when switching to SINFONI-based, Brackett-$\gamma$ astrometry (\cite[Gillessen et al. 2013b]{gil13b}). This orbit is to be preferred because it is less prone to biases. We cannot exclude currently that there might be even a source-intrinsic bias, i.e. that the dust emission of G2 does not exactly trace the gas. That would be an interesting finding in its own.

\begin{figure}
% \vspace*{-2.0 cm}
\begin{center}
 \includegraphics[width=13.5cm]{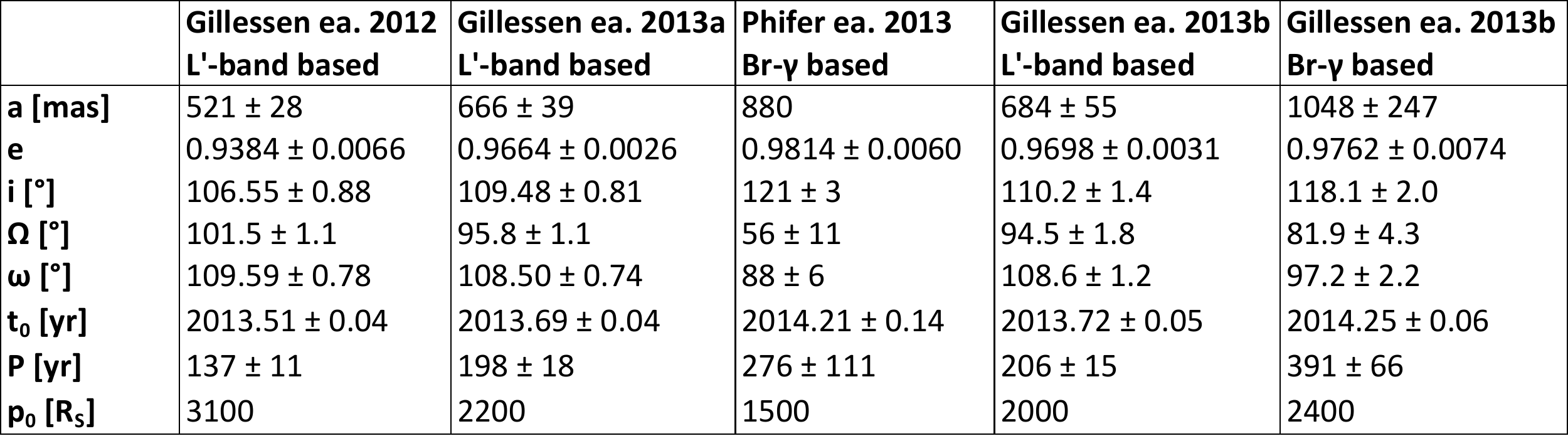} 
% \vspace*{-1.0 cm}
 \caption{Compilation and comparison of the different orbits for G2 published so far. }
   \label{fig1}
\end{center}
\end{figure}

Now, the VLT- and Keck orbit estimates are in good agreement (see figure~\ref{fig1}). The orientation of the orbit is near to that of the inner part of the clockwise stellar disk (\cite[Paumard et al. 2006]{pau06}, \cite[Lu et al. 2009]{lu09}, \cite[Bartko et al. 2009]{bar09}). The estimated pericenter distance is around $r_p \approx 2000 R_S$ (Schwarzschild radii), comparable to the pericenter distance of the famous star S2 on a 16-year orbit (\cite[Gillessen et al. 2009]{gil09}, $1400\,R_S$). The eccentricity is very high with $e\approx 0.98$, which puts strong constraints on the nature of G2. There is a remaining uncertainty on $e$, which is mainly owed to the difficulty of measuring the radial velocity. G2 shows a velocity gradient across the source, and a large intrinsic line width. The latter has grown dramatically in the last few years to around $600\,$km/s, hampering the measurement of the line position. This continued disruption might also mean that the 2013 data are the last trustworthy, and that the true orbit will not be known any better anymore.

\section{Properties of G2}

From the Brackett-$\gamma$ emission we find that G2 is marginally spatially resolved in our SINFONI data. We find an intrinsic Gaussian FWHM size of $42\pm10\,$mas. The fact that we hardly can resolve G2 spatially, while showing a large $v_\mathrm{LSR}$ gradient across the source, is due to the orientation of the orbit. G2 mostly moves along the line-of-sight away from the observer.

The dust emission of G2 can be described by a black-body emission of $T\approx 550\,$K. This relies on the L-band flux, an M-band detection and the limits in K-band (\cite[Gillessen et al. 2012]{gil12}, \cite[Phifer et al. 2013]{phi13}). The finite size of $r \approx 150\,$AU together with the temperature estimate and the observed brightness excludes that the emission comes from an optically thick surface, which would need to have a radius of $r\approx 0.5\,$AU only. 

Using the Brackett-$\gamma$ luminosity of $L\approx 2\times 10^{-3}\, L_\odot$  and
case B recombination yields an estimated mass of G2 of a few times Earth's mass and a characteristic density of $10^5\, \mathrm{cm}^{-3}$. The density is around two orders of magnitude above the ambient density, and G2's mass exceeds the mass in Sgr~A*'s accretion flow enclosed in $r_p$. The total kinetic energy at pericenter passage will be around $10^{45}\,$erg, and if the material falls down to $1\,R_S$, more than $10^{48}\,$erg are available.

Comparing with the UV radiation field in the central arcsecond from the numerous luminous, young, hot stars around Sgr~A*, G2 is plausibly fully ionized. The 
Brackett-$\gamma$ luminosity of G2 has remained constant over the whole time range spanned by spectroscopy from 2004 to 2013. Beyond Brackett-$\gamma$, we also detect G2 in Helium-I ($2.05\,\mu$m) and Paschen-$\alpha$. The latter line happens to be observable between atmospheric absorption bands only due to the high redshift of the emission. The line ratios He-I/Br-$\gamma$ and Pa-$\alpha$/Br-$\gamma$ also remained constant for those epochs, in which we were able to measure them (2008-2013).

\section{Tidal disruption of G2}

\begin{figure}[h]
% \vspace*{-2.0 cm}
\begin{center}
 \includegraphics[width=13.5cm]{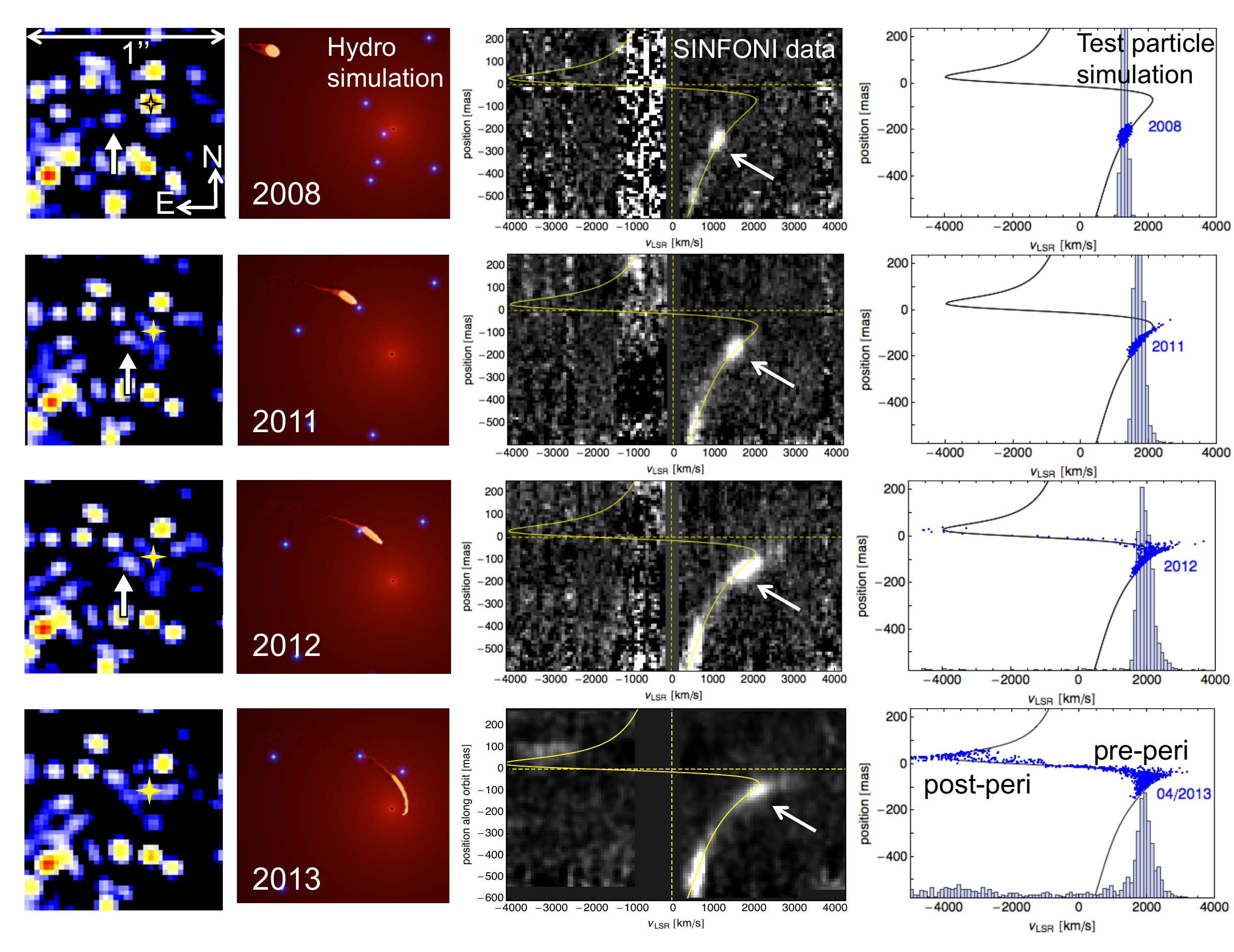} 
% \vspace*{-1.0 cm}
 \caption{From top to bottom: Snapshots in time of the infall of the gas cloud G2 into Sgr~A*. First column: LÕ-band images obtained with NACO of the central arcsecond. The white arrows mark G2, the yellow asterisk Sgr A*. Note that we donÕt detect G2 in 2013. Second column: Snapshots from a hydrodynamic simulation of the infall (\cite[Schartmann et al. 2012]{sch12}). Third column: Position-velocity diagrams from the gas recombination lines of G2 (white arrows) observed with SINFONI, showing the beautiful tidal evolution. Fourth column: A test particle simulation for G2 can describe the 2008-2013 position-velocity diagrams very well.
 }
 \label{fig2}
\end{center}
\end{figure}

A gas cloud with a radius of $150\,$AU in a distance to Sgr~A* of $2800\,$AU (as G2 had in 2008) would need to have a mass of $\approx 10^4\,M_\odot$ in order to be gravitationally bound against the tidal field of the MBH. It is unavoidable, that G2 will undergo tidal disruption, and indeed our data beautifully show the onset and continued evolution of that process. Our SINFONI data are presented best in the form of position-velocity diagrams, where the spatial axis is the line element along the curved trajectory of G2's orbit. Note that it is only possible to perform such an analysis because of the integral field aspect of our spectroscopy. Also, we added up the three diagrams obtained for each epoch from the three emission lines Brackett-$\gamma$, Helium-I, and Paschen-$\alpha$. A few of the resulting diagrams are shown in figure~\ref{fig2}.

\begin{figure}[h]
% \vspace*{-2.0 cm}
\begin{center}
 \includegraphics[width=13.5cm]{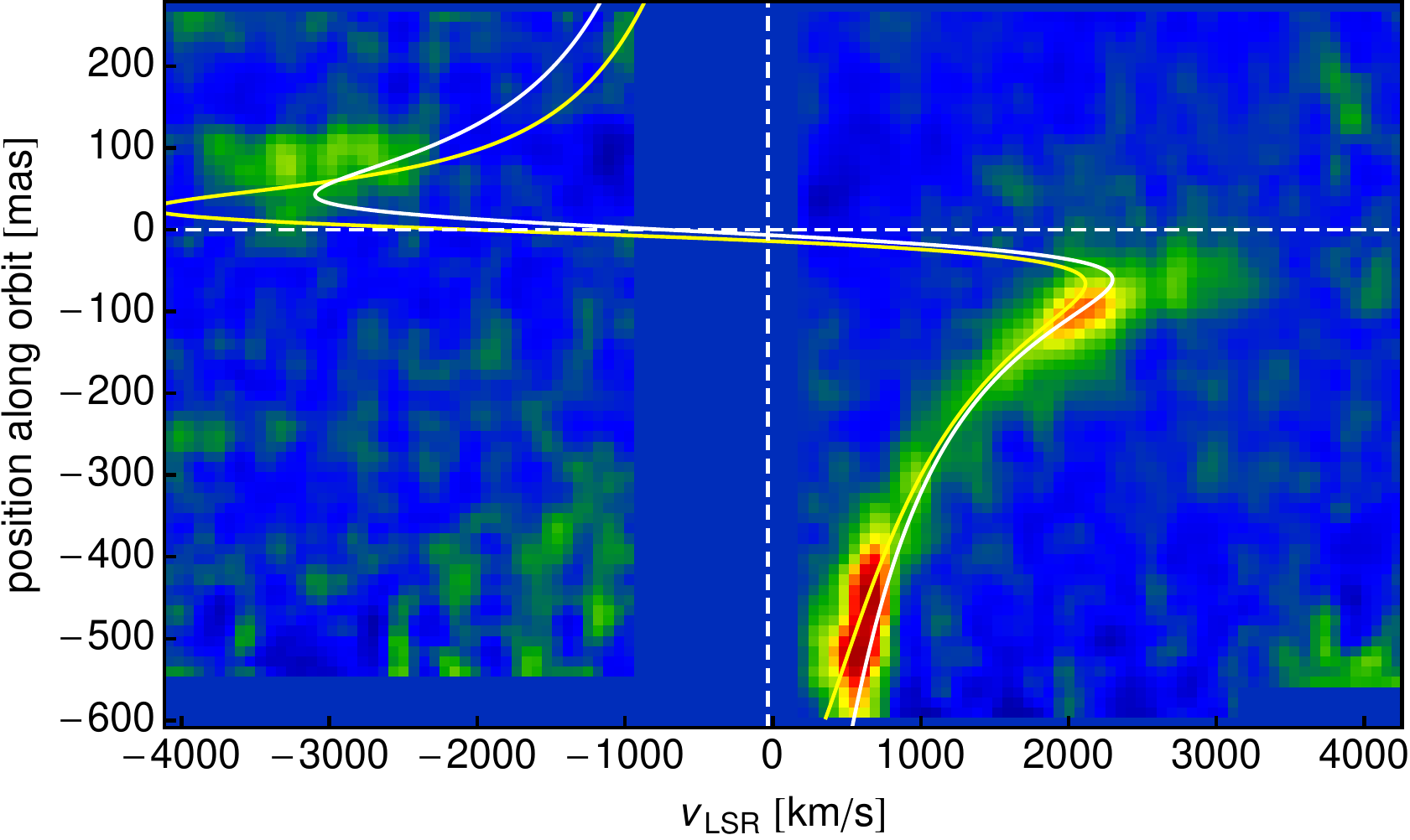} 
% \vspace*{-1.0 cm}
 \caption{Our highest quality position-velocity diagram of G2, extracted from the 2013 April SINFONI data set. The position axis is counted along the orbit projected into the cube (shaded region in the right panel). This diagram is a co-add around the lines Brackett-$\gamma$, Helium-I, and Paschen-$\alpha$ for the red part of the diagram ($v_\mathrm{LSR}>0$), and of the former two lines for the blue side ($v_\mathrm{LSR}<0$). The yellow line delineates the L'-band based orbit, the white line the Brackett-$\gamma$ based one from \cite{gil13b}.
 }
 \label{fig3}
\end{center}
\end{figure}

A simple test particle model can describe the evolution remarkably well. An initially spherical cloud with a Gaussian FWHM of $42\,$mas and
a Gaussian FWHM of the velocity dispersion of $120\,$km/s starting at $t=2000.0$ captures not only the overall evolution of the velocity shear, but also some finer details: 
\begin{itemize}
\item There appears to be gas overshooting in $v_\mathrm{LSR}$ the bulk of the emission in the 2012 data set. This is also seen in the test particle simulation.
\item In the April 2013 data set, some of the gas is already detected at the blue shifted side, after pericenter (figure~\ref{fig3}). Also this is expected from the test particle simulation. 
\end{itemize}
\smallskip
The emission on the blue-shifted side is at the SNR limit of our data set. 

It is worth pointing out, that the test particle simulation also predicts that the pericenter flyby has a significant intrinsic duration of over one year, simply due to the tidal stretching. In this sense, there does not exist a well-defined pericenter date.

In the position-velocity diagrams there seems to be a tail of gas following the same orbit as G2. The whole structure might thus be a much more elongated gas feature, of which G2 appears to be the head. \cite{phi13} have questioned the physical connection of tail and head. The lower $v_\mathrm{LSR}$ of the tail means that it could also be a background feature in the general ambient gas. In addition, the tail emission in image space does only roughly follow the orbital trace. Nevertheless, our unparalleled, ultra-deep integral field spectroscopy reveals that there is a fainter 'bridge' of emission between head and tail following exactly the orbit in the position-velocity diagram. It is visible in figure~\ref{fig3}. This argues in favor of a physical connection.

\section{Interactions of G2}

The fact that the tail does not perfectly match the orbit in image space might also point towards an additional effect: It is expected that the gas of G2 interacts with the hot gas of the accretion flow around Sgr~A*. This is actually one of the most exciting aspects of the G2~-Sgr~A* encounter.

Accretion flow models for Sgr~A* have been designed to explain its extremely low (radio) luminosity. These models predict that there is an atmosphere of hot, thin gas around Sgr~A* that extends out to roughly the Bondi radius ($\approx 10^5 R_S$). The density profile of the atmosphere depends on the model type. 
For a radiatively inefficient accretion flow one has $\rho(r) \sim r^{-1}$ (\cite[Yuan et al. 2003]{yua03}), while for an advection dominated accretion flow $\rho(r) \sim r^{-3/2}$ (\cite[Narayan \& Yi 1995]{nar95}), and for a convection dominated accretion flow the profile can be as flat as $\rho(r) \sim r^{-1/2}$ (\cite[Quataert \& Gruzinov 2000]{qua00}).

The observational constraints on the density profile are rather weak: The diffuse X-ray emission around Sgr~A* as resolved by Chandra is due to the accretion flow (\cite[Baganoff et al. 2003]{bag03}, \cite[Wang et al. 2013]{wan03}) and yields a constraint roughly at the Bondi radius. The rotation measure of Sgr~A* obtained in the submm domain on the other hand allows estimating the density in the innermost $10 R_S$ (\cite[Marrone et al. 2007]{mar07}).

G2 now might offer the unique opportunity to probe the accretion flow on scales from $10^4 R_S$ to $10^3 R_S$. \cite{gil12} estimated that detectable X-ray radiation might occur during the pericenter approach due to a shock front developing (\cite[McKee \& Cowie 1975]{mck75}). The two main unknowns for that are the volume filling factor of G2, and the density profile of the ambient gas. Predicting the full evolution of G2, however, is a more complex problem, because the hydrodynamic time scales all are around a few years. In particular, the Kelvin-Helmholtz and Rayleigh-Taylor instabilities might be important. Hence, hydrodynamic simulations are needed, and have been performed (\cite[Burkert et al. 2012]{bur12}, \cite[Schartmann et al. 2012]{sch12}, \cite[Anninos et al. 2012]{ann12}, \cite[Abarca et al. 2013]{aba13}).

The simulations consistently show that the evolution of G2 up to pericenter is very similar to what one expects from tidal shearing only. After pericenter passage, the tidal description breaks down and hydrodynamic effects start to dominate the evolution. The details depend strongly on the assumptions made about G2 and the ambient gas. Turning the argument around, the combination of observing the line profile evolution and possible radiative reactions can be used to constrain the assumptions put into the simulations, and thus to learn about the structure of the accretion flow around the prototypical low-luminosity AGN, Sgr~A*.

\section{The nature of G2}

The nature of G2 is currently being discussed. While it was called 'gas cloud' in the discovery paper (\cite[Gillessen et al. 2012]{gil12}), other papers have proposed that actually there resides a (faint and undetected) star at the origin of G2. It is interesting to review the stellar and the gas cloud model in the following. 

The original proposal of G2 being a gas cloud had a serious shortcoming. The orbit of G2 lies in the plane of the clockwise stellar disk and the apocenter as calculated from the orbit presented in \cite{gil12} is at the inner edge of that disk at $r\approx 1''$. Hence, a connection is likely, for example as a collisional product of stellar winds in the region. The shortcoming is that if G2 formed there in pressure equilibrium and started falling inward then, it should have been already tidally stretched into an almost linear feature by the time we detected it. This is not what is seen in the 2004 - 2008 data, where G2 appears rather compact. On that orbit, G2 must have formed relatively recently, somewhere in the years 1990 - 2000, and thus in a radial range well inside the inner edge of the stellar disk. 

The updated orbit now has a larger apocenter distance of $r\approx 2''$. That opens a new possibility: G2 could have formed half way in on the orbit and still would originate from the disk. Giving it a significant inward 'birth' velocity (such as to place it on the observed orbit) solves the issue of too early tidal disruption. Hence, on that orbit G2 can have formed in the stellar disk and remained compact most of the way in. This lifts the most serious shortcoming of the pure gas cloud model. 

\cite{shc13} introduces the idea of a magnetically arrested cloud, takes into account the explicit positions of individual stars as ionizing sources, and models the dust emission. The model can explain the absolute values of both the line luminosities and the L-band and M-band magnitudes, as well as the velocity dispersions. It probably currently is the best 'pure gas cloud' model. 

In the gas cloud scenario, G2 could have formed from collisions of stellar winds. Such clumps, with few Earth masses, have already been seen in simulations of the stellar winds around Sgr~A* (\cite[Cuadra et al. 2006]{cua06}), although these particular simulations were performed with an SPH-code, which is not well suited for the problem at hand. Follow-up work seems to indicate that indeed at any moment in time there are a several ten clumps that resemble G2 within the simulated volume of $r\approx10''$. 
G2 might thus be special only in the sense that it is on a very radial orbit. L-band images of the GC region show dusty sources all over, but the detection of G2 as a gas cloud was only possible in the central arcsecond, since there $v_\mathrm{LSR}$ is large enough to Doppler shift the emission away from the general mini spiral emission. 

It is also worth noting that one can pinpoint the most likely progenitor stars for that scenario: First, there is the eclipsing binary IRS16SW (\cite[Martins et al. 2006]{mar06}), which consists of two $50 M_\odot$ stars, and secondly the massive, young star S91. Both objects are part of the clockwise stellar disk, and their position on the disk is consistent with being the origin of G2. 

Soon after the discovery of G2, \cite{mur12} proposed a completely different model for G2: It might be the evaporating protoplanetary disk around a fainter star. Such a disk is not tidally stable at the position of G2, and hence it gets disrupted, the closer the object gets to the MBH. The gas of the disk is then ionized by the UV radiation field of the surrounding stars and the recombination lines can be observed. One could call such an object a 'tidal comet'. This model is attractive mainly for two reasons: It can naturally explain the compactness of G2. And a priori, the presence of stars is more likely close to Sgr~A*, since gas clouds are short-lived. The implications of that model are interesting: The debris around a star would flag the star, which itself could be too faint to be observable. And one would need to speculate about planet formation in the GC.

This model also has two difficulties. The high eccentricity means that one needs a rather strong kick for the star to change its orbit from one, which does not disrupt the protoplanetary disk at pericenter passage, to the current one. Yet, the kick itself should not destroy the protoplanetary disk either. The authors estimate how likely that is, and conclude that "we are somewhat fortunate to observe" G2. Secondly, the observed constancy of the line luminosities does not match what one would expect for that model. Coming closer to the MBH, more and more mass of the protoplanetary disk is lifted, leading to an increase in luminosity. 

Also the model of \cite{sco13} places a star at the origin of G2. They propose that the relatively slow wind of a T-Tauri star would create a shock front, the emission of which we see as G2. For this model, a scattering event to the high eccentricity orbit is unproblematic. Also the constancy of the luminosity is easier to reconcile with this model. The authors investigate the various sources of ionization and recombination. The inner, cold shock dominates the emission and is collisionally ionised from the stellar wind. In this model, the luminosity thus depends on the wind parameters and is independent of the orbit.

The explicit hydrodynamic simulations of \cite{bal13} for a stellar wind source plunging through the accretion flow of Sgr~A*, however, show that in the radial range through which G2 was observed, an increase of the luminosity by a factor of a few would still be expected - which renders thus the T-Tauri star model less likely. Overall, the constancy of the line emission appears to be difficult to explain in any stellar scenario.

The observational work by \cite{phi13} favors a stellar origin of G2, based, however, not on their actual data, but expressing the prior that it should be more likely to find stars
around Sgr~A* than gas clouds. Their highest quality OSIRIS data set from 2006 confirms that G2 is an extended gas cloud, and is the earliest data set resolving G2. Later OSIRIS observations are mainly SNR-limited. The paper also presents the best upper limit on a potential K-band source at the position of G2: mag$_{K'} > 20$. 

This limit is at odds with the results presented in \cite{eck13}. These authors claim to find a K-band counterpart of G2 in NACO/VLT-based images, and presented during this conference similar claims from a Keck data set. Given that the two other obervational groups did not see any significant evidence for a K-band counterpart from de facto identical data sets, the findings of \cite{eck13} are unexpected. 

But even if a K-band source can be identified, one cannot firmly conclude on the existence of a stellar source inside G2. As shown by \cite{gil12} and \cite{eck13}, a dust temperature of $550\,$K matches the L-band and M-band fluxes, and yields a K-band flux that would be compatible with the putative detection. \cite{eck13} note that a slightly cooler dust temperature ($450\,$K) together with an embedded, faint star can also reproduce the photometry. In other words, a K-band detection is still ambiguous. Only if one detected G2 in H-band, one would be able to conclude that G2 contains a star. So far, only upper limits have been reported for H-band.

The currently observed, beautiful tidal evolution of the gas is only weakly dependent on whether one places a star inside of G2 or not. At the observed distance from Sgr~A*, the tidal force of the MBH dominates over the gravity of the supposed star inside of G2. Differences only occur when the embedded object replenishes material. Then the resulting
density profile for G2 is more centrally concentrated compared to a pure gas cloud. This in turn, leads to a different velocity gradient evolution in the position-velocity diagram (\cite[Gillessen et al. 2013b]{gil13b}). Comparing with the observed evolution, a pure gas cloud model seems to be the better match currently, but the parameter space for possible stellar scenarios certainly has not yet been systematically checked.

More models have been proposed for G2: \cite{sch12} noted that the head-tail geometry of G2 could be explained by an (originally) ring-like geometry of G2. \cite{mey12} propose that a nova is at the origin of G2. Both ideas suffer from the fact that in order to match the observations, one needs to finetune the moment when G2 started to expand. 
\cite{mir12} suggested that the debris of a collision between a low-mass star and a stellar black hole could create G2. The debris could have been on the same orbit for hundreds of revolutions, avoiding the finetuning problem. 

The further evolution of G2 probably will tell about its true nature. The different models might vary dramatically on what will happen to G2 during and after pericenter passage. A pure gas cloud model predicts a complete disruption of G2, while the stellar models predict that the gas cloud should reform after pericenter passage. Observations during and after pericenter passage will probably shed decisive light on the nature of G2.

\section{The future of G2}

While it is clear that the gas currently observed cannot survive as compact cloud the upcoming pericenter passage, it is less clear what observable consequences the fly-by might have. Quite a number of observing proposals\footnote{For an overview of 2013 and 2014 G2 related observing proposals, see:
\texttt{https://wiki.mpe.mpg.de/gascloud/ProposalList}. For additions to the list, please contact the authors.
}
have been focusing on the G2 event. So far, no observation of any G2 related radiation increase has been reported in any waveband. Nevertheless, it is interesting to
look at the various ideas.

The first observable sign of G2 plunging through the accretion flow might be from the shock front that is expected to form. \cite{gil12} estimated the amount of X-ray radiation originating from the shock-heated gas. The temperature might reach up to $10^7\,$K, and the luminosity in the observable $2-8\,$keV band would exceed by a small factor the current quiescent level of emission from Sgr~A*. 

Shock accelerations of electrons in the bow shock might lead to radio emission (\cite[Narayan, \"Ozel \& Sironi 2012]{nar12}, \cite[Sadowski et al. 2013a]{sad13a}, \cite[Crumley \& Kumar 2013]{cru13}). In the $0.1\,$GHz to $1\,$GHz band, the emission can be much higher than the source intrinsic radio flux. \cite{sad13b} showed that the peak of the radio emission might occur even nine months before the nominal pericenter date. The VLA is conducting currently a public monitoring porgram of Sgr~A*\footnote{For details see: \texttt{https://science.nrao.edu/enews/5.10/index.shtml\#g2\_encounter}}, and so far no flux increase has been reported (but see the multi-year radio light curve of Sgr~A* in \cite[Beaklini \& Abraham 2013]{bea13}).

Sgr~A* is expected to host a cusp of stellar mass black holes around it (\cite[Morris 1993]{mor93}). \cite{bar13} study the possibility that the G2-cloud collides with such an object. They conclude that under favorable circumstances the event might be detectable. Close to pericenter passage, multiple encounters are likely to occur.

The rotation measure of Sgr~A*, a measure for the amount of material between the source and the observer, has been stable for many years (\cite[Marrone et al. 2007]{mar07}). Additional gas coming from G2 could manifest itself thus in a change of the quantity. A change occurring in the near future would be very likely to be associated with gas of G2 arriving at Sgr~A*.

More speculative are the ideas, what happens if material enters the innermost accretion zone. It is clear, that this might take much longer than the actual fly-by, since the orbit only deposits material at around $2000\,R_S$, and significant radiation is produced in the central $10\,R_S$. The spiral down is probably dominated by the viscous time scale, such that the process might even take years (\cite[Moscibrodzka et al. 2012]{mos12}). An increase in the accretion rate onto Sgr~A* would lead to increased emission across all wavebands (\cite[Yuan et al. 2004]{yua04}). 
This is different from the mechanism believed to create the NIR and X-ray flares of Sgr~A* (\cite[Genzel et al. 2003]{gen03}, \cite[Baganoff et al. 2001]{bag01}). They are due to local heating of synchrotron emitting electrons (\cite[Dodds-Eden et al. 2009]{dod09}), rather than a global change in accretion rate. Yet, additional material in the accretion flow might change the flaring characteristics - flares could happen more often, last longer, or shine brighter than currently (\cite[Dodds-Eden et al. 2011]{dod11}).

Even more speculative are ideas that Sgr~A* could change its overall accretion state. Currently, the accretion flow is optically thin and geometrically thick. At higher rates, it might change to an optically thick, geometrically thin state, i.e. develop a pronounced accretion disk. In addition, Sgr~A* might exhibit visible jets then. Such structural changes might be observable in two ways: While the resolved mm-VLBI measurements (\cite[Doeleman et al. 2008]{doe08}) do not yet allow reconstructing an image of Sgr~A*, the observed visibilities could show structural changes. The other way to detect these might be astrometry. The position of Sgr~A* can be determined to $\approx 100\,\mu$as (\cite[Reid et al. 2008]{rei08}). The expected change in effective photocenter position of Sgr~A* due to jets developing might exceed that number, and thus VLBI astrometry has the power of detecting such a change.

Maybe it is not even surprising to observe variations in the accretion rate of Sgr~A*. The work of \cite{cua06} showed that it is expected to vary significantly on longer time scales, and scaling to shorter time scales then would predict still some level of variation. There are also observational hints for it: From the ISM surrounding Sgr~A* one can detect X-ray reflection radiation (\cite[Muno et al. 2007]{mun07}, \cite[Clavel et al. 2013]{cla13}). The emission moves outward, away from Sgr~A*, and is most likely a light echo. If so, there must have been a source that reached $10^{39}\,$erg/s a few hundred years ago in the GC, Sgr~A* being the most likely candidate.

\section{Final remarks}
The discovery of G2 has triggered large interest, among observers and theorists, as well as from the general public. It might constitute the unique opportunity of observing in real-time a MBH being fed, a process which takes place throughout cosmic time and the universe. G2 might deliver unexpected insights into MBH growth or 
accretion flows around MBHs. Continued observing might be highly rewarding, and even non-detections can be very telling. We wish to encourage further observations - and reporting thereof. Be it detections or non-detections.

\end{document}